\documentclass[conference]{IEEEtran}

\usepackage{ctable}
\usepackage{cite}
\usepackage[cmex10]{amsmath}
 \usepackage{acronym}
\usepackage{graphicx}
\usepackage{multirow}
\usepackage{listings}
\usepackage{color}
\usepackage{xcolor}
\usepackage{balance}

\colorlet{@punct}{red!60!black}
\definecolor{@delim}{RGB}{20,105,176}

\lstdefinelanguage{json}{
    basicstyle=\footnotesize\ttfamily,
    literate=
     *{\ }{{{\ }}}{1}
      {:}{{{\color{@punct}{:}}}}{1}
      {,}{{{\color{@punct}{,}}}}{1}
      {\{}{{{\color{@delim}{\{}}}}{1}
      {\}}{{{\color{@delim}{\}}}}}{1}
      {[}{{{\color{@delim}{[}}}}{1}
      {]}{{{\color{@delim}{]}}}}{1},
}

\newcommand{\includeJSON}[1]{\lstinputlisting[language=json,firstnumber=1]{#1}}

\acrodef{KWAPI}{KiloWatt API}
\acrodef{PUE}{Power Usage Effectiveness}
\acrodef{IPMI}{Intelligent Platform Management Interface}
\acrodef{PDU}{Power Distribution Unit}
\acrodef{ePDU}{enclosure PDU}
\acrodef{JSON}{JavaScript Object Notation}
\acrodef{RRD}{Round-Robin Database}

\begin{document}

\title{A Generic and Extensible Framework for Monitoring Energy Consumption of OpenStack Clouds}

\author{\IEEEauthorblockN{Fran\c{c}ois Rossigneux, Jean-Patrick Gelas, Laurent Lef\`{e}vre, Marcos Dias de Assun\c{c}\~ao}
\IEEEauthorblockA{Inria Avalon, LIP Laboratory\\
Ecole Normale Sup\'{e}rieure de Lyon\\
University of Lyon, France}
}

\maketitle

\begin{abstract}
Although cloud computing has been transformational to the IT industry, it is built on large data centres that often consume massive amounts of electrical power. Efforts have been made to reduce the energy clouds consume, with certain data centres now approaching a \ac{PUE} factor of 1.08. While this is an incredible mark, it also means that the IT infrastructure accounts for a large part of the power consumed by a data centre. Hence, means to monitor and analyse how energy is spent have never been so crucial. Such monitoring is required not only for understanding how power is consumed, but also for assessing the impact of energy management policies. In this article, we draw lessons from experience on monitoring large-scale systems and introduce an energy monitoring software framework called \ac{KWAPI}, able to handle OpenStack clouds. The framework --- whose architecture is scalable, extensible, and completely integrated into OpenStack --- supports several wattmeter devices, multiple measurement formats, and minimises communication overhead.
\end{abstract}

\IEEEpeerreviewmaketitle

\section{Introduction}
\acresetall

Cloud computing \cite{ArmbrustCloud:2009} has become a key building block in providing IT resources and services to organisations of all sizes. Among the claimed benefits of clouds, the most appealing derive from economies of scale and often include a pay-as-you-go business model, resource consolidation, elasticity, good availability, and wide geographical coverage. Despite these advantages when compared to other provisioning models, in order to serve customers with the resources and elasticity they need, clouds generally rely on large data centres that consume massive amounts of electrical power \cite{BaligaInternet:2011}\cite{GreenbergCostCloud:2009}.
 
Although some data centres now approach a \ac{PUE} factor of 1.08\footnote{http://gigaom.com/2012/03/26/whose-data-centers-are-more-efficient-facebooks-or-googles/}, such a mark means that the IT infrastructure is now responsible for a large part of the consumed power. Means to monitor and analyse how energy is spent are crucial to further improvement, but our previous work in this area has demonstrated that monitoring the power consumed by large systems is not always an easy task \cite{OrgerieSaveWatts:2008,AssuncaoIngrid:2010,DaCostaGreenNet:2010}. There are multiple power probes available in the market, generally with their own APIs, physical connections, precision, and communication protocols\cite{eelsd2013}. Moreover, cost related constraints can lead data centre operators to acquire and deploy equipments at multiple stages, or to monitor the power consumption of only part of an infrastructure.

From a cost perspective, monitoring the power consumption of only a small part of deployed equipments is sound, but it prevents one from capturing important nuances of the infrastructure. Previous work has shown that as a computer cluster ages, certain components wear out, while others are replaced, leading to heterogeneous power consumption among nodes that were seemingly homogeneous \cite{DeanGoogleFailures:2008}. The difference between nodes that consume the least power and nodes that consume the most can reach 20\% \cite{MehdiHeterogeneous:2013}, which reinforces the idea that monitoring the consumption of all equipments is required for exploring further improvement in energy efficiency and evaluate the impact of system-wide policies. Monitoring a great number of nodes, however, requires the design of an efficient infrastructure for collecting and processing the power consumption data.

This paper describes the design and architecture of a generic and flexible framework, termed as \ac{KWAPI}, that interfaces with OpenStack to provide it with power consumption information collected from multiple heterogeneous probes. OpenStack is a project that aims to provide ubiquitous open source cloud computing platform and is currently used by many corporations, researchers and global data centres\footnote{http://www.openstack.org/user-stories/}. We describe how \ac{KWAPI} is integrated into Ceilometer; OpenStack's  component conceived to provide a framework to collect a large range of metrics for metering purposes\footnote{https://wiki.openstack.org/wiki/Ceilometer}. With the increasing use of Ceilometer as the \textit{de facto} metering tool for OpenStack, we believe that such an integration of a power monitoring framework into OpenStack can be of great value to the research community and practitioners.

The remaining part of this paper is organised as follows. Section~\ref{sec:related_work} describes background and related work, whereas Section~\ref{sec:architecture} presents the \ac{KWAPI} architecture. Section~\ref{sec:performance} discusses experimental results on measuring the throughput of \ac{KWAPI}, and Section~\ref{sec:conclusion} concludes the paper.


\section{Background and Related Work}
\label{sec:related_work}

This section provides an overview of Ceilometer's architecture and describes related work on monitoring power consumption of large-scale computing infrastructure.

\subsection{OpenStack Ceilometer}

\begin{figure}[!htb]
\center
\includegraphics[width=1.\columnwidth]{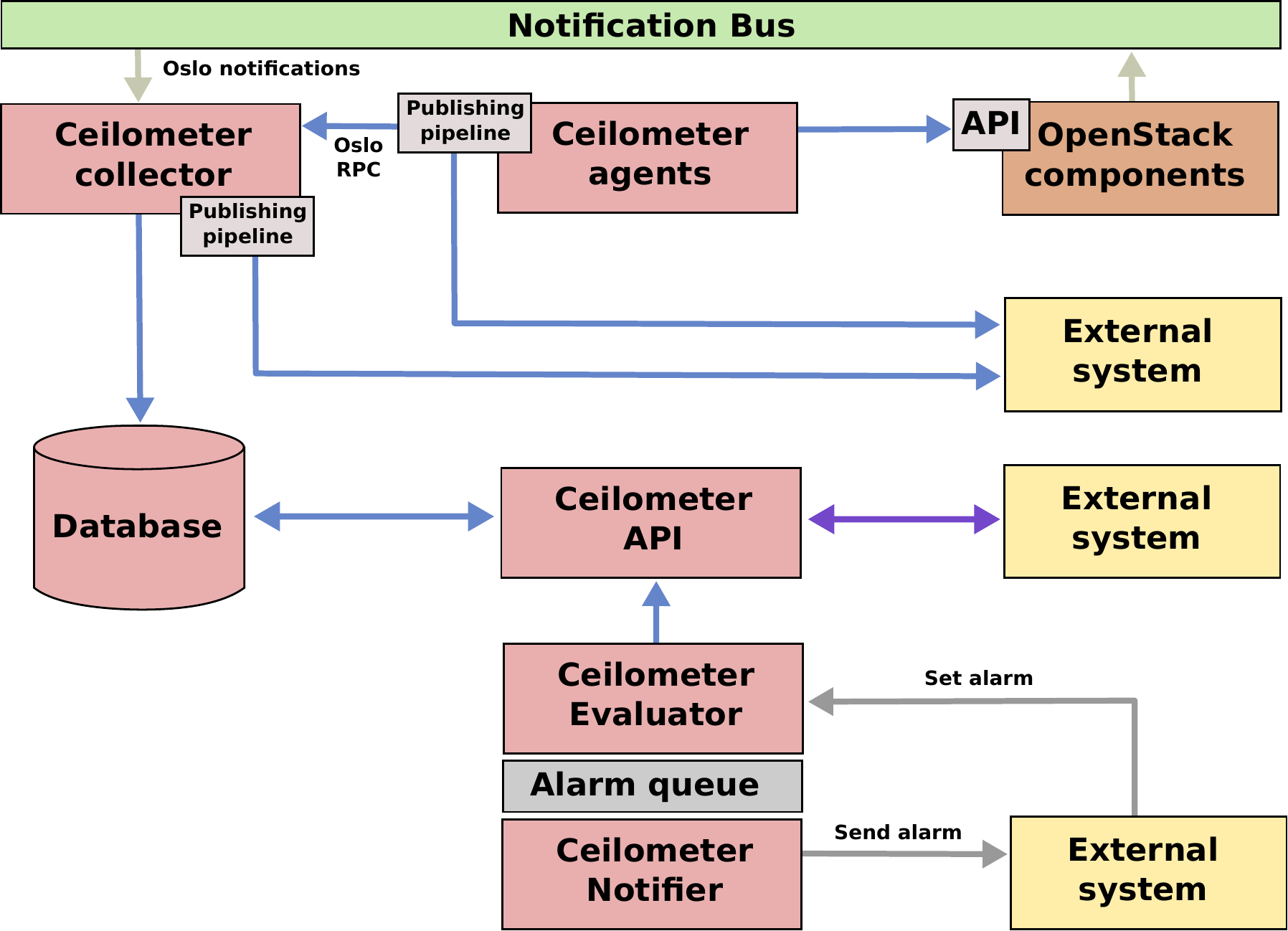}
\caption{Overview of Ceilometer's logical architecture.}
\label{fig:arch_ceilometer}
\end{figure}

Ceilometer --- whose logical architecture\footnote{http://docs.openstack.org/developer/ceilometer/architecture.html} is depicted in Figure~\ref{fig:arch_ceilometer} --- is OpenStack's framework for collecting performance metrics and information on resource consumption. As of writing, it allows for data collection under three methods:

\begin{itemize}
\item \textbf{Bus listener agent}, which picks events on OpenStack's notification bus and turns them into Ceilometer samples (\textit{e.g.} cumulative type, gauge or delta) that can then be stored into the database or provided to an external system via publishing pipeline.

\item \textbf{Push agents}, more intrusive, consist in deploying agents on the monitored nodes to push data remotely to be taken by the collector.

\item \textbf{Polling agents} that poll APIs or other tools to collect information about monitored resources.
\end{itemize} 

The last two methods depend on a combination of central agent, computer agents and collector. The compute agents run on nodes and retrieve information about resource usage related to a given virtual machine instance and a resource owner. The central agent, on the other hand, executes \textit{pollsters} on the management server to retrieve data that is not linked to a particular instance. Pollsters are software components executed, for example, to poll resources by using an API or other methods. The Ceilometer database, which can be queried via Ceilometer API, allows an external system to view the history of a resource's metrics, and enables the system to set and receive alarms.

The \textit{hmac} module of Python's library can be used for signing metering messages, and a shared secret value can be provided in the configuration settings. The collector and systems accessing the API use signatures included in the messages for verification.

\subsection{Energy Monitoring and Efficiency in Clouds}

Over the past years, several techniques have been provided to minimise the energy consumed by computing infrastructure. At the hardware level, for instance, processors are able to operate at multiple frequency and voltage levels, and the operating systems or resource managers can choose the level that matches the current workload \cite{LaszewskiDVFS:2009}. At the resource management level, several approaches are proposed, including resource consolidation \cite{BeloglazovOpenStack:2014} and rescheduling requests \cite{OrgerieSaveWatts:2008}, generally with the goal of switching off unused resources or setting them to low power consumption modes. Attempts have also been made to assess the power consumed by individual applications \cite{NoureddineThesis:2014}.

A means to monitor the energy consumption is key to assess potential gains of techniques to improve software and cloud resource management systems. Cloud monitoring is not a new topic \cite{AcetoMonitoring:2013} as tools to monitor computing infrastructure \cite{BrinkmannMonitoring:2013,VarretteICPP:2014} as well as ways to address some of the usual issues of management systems have been introduced \cite{WardMonitoring:2013,TanMonitoring:2013}. Moreover, several systems for measuring the power consumed by compute clusters have been described in the literature \cite{AssuncaoIngrid:2010}. As traditional system and network monitoring techniques lack the capability to interface with wattmeters, most approaches for measuring energy consumption have been tailored to the needs of projects for which they were conceived.

In our work, we draw lessons from previous approaches to monitor and analyse energy consumption of large-scale distributed systems \cite{OrgerieSaveWatts:2008,DaCostaGreenNet:2010,AssuncaoIngrid:2010,MehdiHeterogeneous:2013,CGC2012}. We opt for creating a framework and integrating it with a successful cloud platform (\textit{i.e.} OpenStack), which we believe is of value to the research community and practitioners working on the topic.


\section{The \ac{KWAPI} Architecture}
\label{sec:architecture}

\begin{figure*}[!htb]
\center
\includegraphics[width=0.7\linewidth]{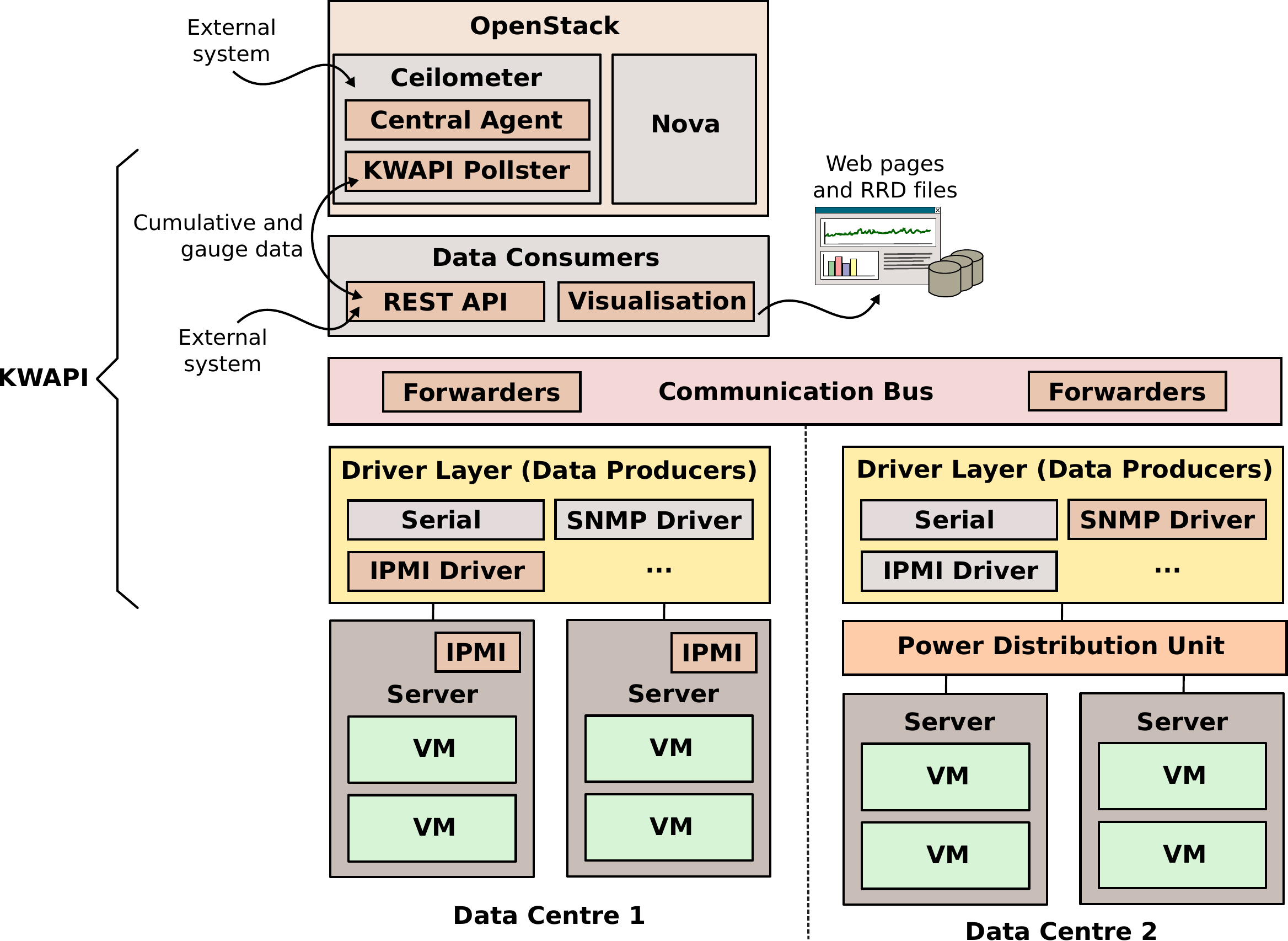}
\caption{Overview of \ac{KWAPI}'s architecture.}
\label{fig:architecture}
\end{figure*}

An overview of the \ac{KWAPI} architecture is presented in Figure~\ref{fig:architecture}. The architecture follows a publish/subscribe model based on a set of layers comprising:

\begin{itemize} 
\item \textbf{Drivers}, considered data producers responsible for measuring the power consumption of monitored resources and providing the collected data to consumers via a communication bus; and 
\item \textbf{Data Consumers} --- or \textbf{Consumers} for short --- that subscribe to receive and process the measurement information. 
\end{itemize}

The communication between layers is handled by a bus, as explained in detail later. Data consumers can subscribe to receive information collected by drivers from multiple sites. Both drivers and consumers are easily extensible to support, respectively, several types of wattmeters and provide additional data processing services. A REST API is designed as a data consumer to provide a programming interface for developers and system administrators. In this work it is used to interface with OpenStack by providing the information (\textit{i.e.} by polling monitored devices) required by a \textit{\ac{KWAPI} Pollster} to feed Ceilometer.

The following sections provide more details on the main architecture components and their relationship with OpenStack Ceilometer.

\subsection{Driver Layer}

Drivers are threads initialised by a Driver Manager with a set of parameters loaded from a file compliant with the OpenStack configuration format. These parameters are used to query the meters (\textit{e.g.} IP address and port) and determine the sensor ID to be used in the collected metrics. The measurements that a driver obtains are represented as \ac{JSON} dictionaries that maintain a small footprint and that can be easily parsed. The size of dictionaries varies depending on the number of fields set by drivers (\textit{i.e.} whether message signing is enabled). 

Figure~\ref{fig:json} shows a simple example of a \ac{JSON} payload containing one measurement. Optional fields such as voltage and current can be included. ACK messages have a fixed size of 66 bytes when using TCP connection; drivers and data consumers communicate via IPC sockets when running on the same machine.

\begin{figure}
\includeJSON{measurement.json}
\caption{Example of \ac{JSON} payload.}
\label{fig:json}
\end{figure}

Drivers can manage incidents themselves, but the manager also checks periodically if all threads are active, restarting them if necessary. It is important to avoid losing measurements because the reported information is in W instead of kWh. The loss of a measurement may be significant. 

Wattmeters available in the market vary in terms of physical interconnection, communication protocols, packaging and precision of measurements they take. They are mostly packaged in multiple outlet power strips called \acp{PDU} or \acp{ePDU}, and more recently in the \ac{IPMI} cards embedded in the computers themselves. Support for several types of wattmeter has been implemented, which drivers can use to interface with a wide range of equipments. In our work, we used \ac{IPMI} initially at Nova to shutdown and turn on compute nodes, but nowadays we also use it to query a computer chassis remotely.

Although Ethernet is generally used to transport \ac{IPMI} or SNMP packets over IP, USB and RS-232 serial links are also common. Wattmeters that use Ethernet are generally connected to an administration network (isolated from the data centre main data network). Moreover, wattmeters may differ in the manner they operate; some equipments send measurements to a management node on a regularly basis (push mode), whereas others respond to queries (pull mode). Other characteristics that differ across wattmeters include: 

\begin{itemize}
\item refresh rate (\textit{i.e.} maximum number of measurements per second);
\item measurement precision; and 
\item methodology applied to each measurement (\textit{e.g.} mean of several measurements, instantaneous values, and exponential moving averages).
\end{itemize}

Table \ref{tab:wattmeters} shows the characteristics of equipments we deployed and used with Kwapi in our cloud infrastructure.

\begin{table}
\centering
\caption{Wattmeter infrastructure}
\label{tab:wattmeters}
\begin{footnotesize}
\begin{tabular}{llcc}
\toprule
\multirow{2}{18mm}{\textbf{Device Name}} & \multirow{2}{30mm}{\textbf{Interface}} & \multirow{2}{12mm}{\centering{\textbf{Refresh Time (s)}}} & \multirow{2}{10mm}{\centering{\textbf{Precision (W)}}}  \\
& & & \\
\toprule
Dell iDrac6    & IPMI / Ethernet           & 5    & 7 \\
\midrule
Eaton          & Serial, SNMP via Ethernet & 5    & 1 \\
\midrule
OmegaWatt      & IrDA Serial               & 1    & 0.125 \\
\midrule
Schleifenbauer & SNMP via Ethernet         & 3    & 0.1 \\
\midrule
Watts Up?      & Proprietary via USB       & 1    & 0.1 \\
\midrule
ZEZ LMG450     & Serial                    & 0.05 & 0.01 \\
\bottomrule
\end{tabular}
\end{footnotesize}
\end{table}

\subsection{Data Consumers}

A data consumer retrieves and processes measurements taken by drivers and provided via bus. Consumers expose the information to other services including Ceilometer and visualisation tools. By using a system of prefixes, consumers can subscribe to all producers or a subset of them. When receiving a message, a consumer verifies the signature, extracts the content and processes the data. By default \ac{KWAPI} provides two data consumers, namely the REST API (used to interface with Ceilometer) and a visualisation consumer.

\subsubsection{REST API}

The API consumer computes the number of kWh of each driver probe, adds a timestamp, and stores the last value in watts. If a driver has not provided measurements for a long time, the corresponding data is removed. The REST API allows an external system to retrieve the name of probes, measurements in W or kWh, and timestamps. The API is secured by OpenStack Keystone tokens\footnote{http://keystone.openstack.org}, whereby the consumer needs to ensure the validity of a token before sending a response to the system. 

\subsubsection{Visualisation}

\begin{figure*}[!htb]
\center
\includegraphics[width=0.9\linewidth]{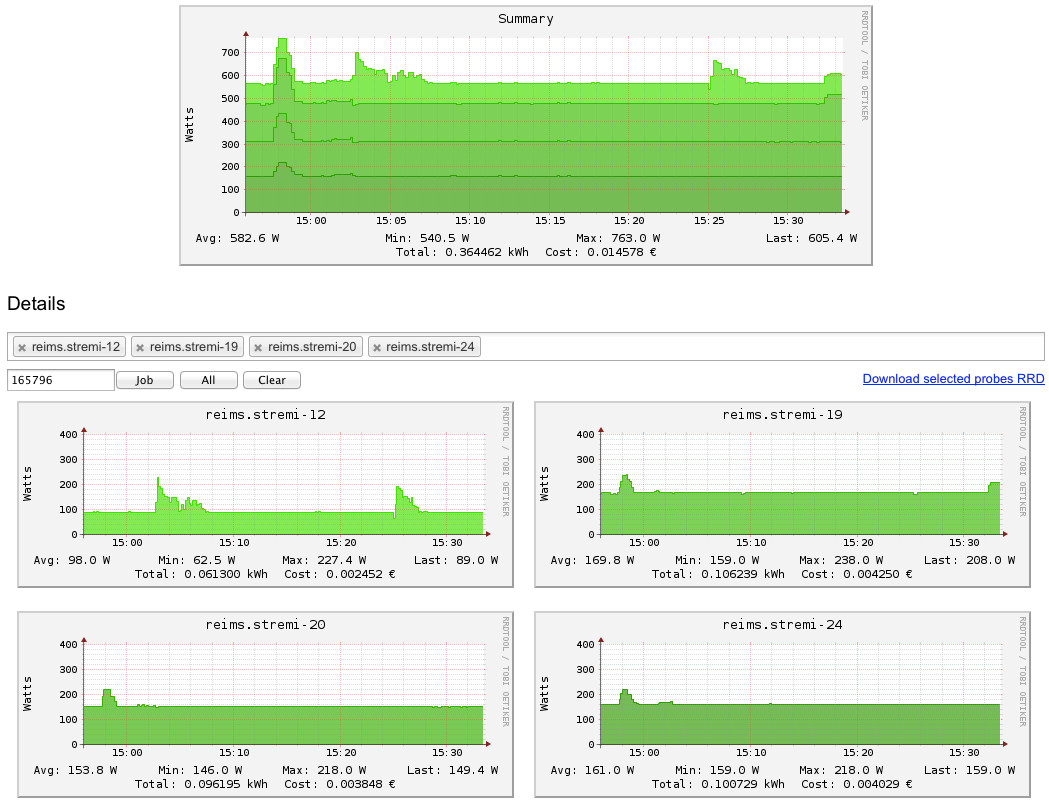}
\caption{Example of graph generated by the visualisation plug-in (4 monitored servers).}
\label{fig:graph_example}
\end{figure*}

The visualisation consumer builds \ac{RRD} files from received measurements, and generates graphs that show the energy consumption over a given period, with additional information such as average electricity consumption, minimum and maximum watt values, last value, total energy and cost in Euros. \ac{RRD} files are of fixed size and store several collections of metrics with different granularities. A web interface displays the generated graphics and a cache mechanism triggers the creation of graphs during queries only if they are out of date. These visualisation resources offer quick feedback to administrators and users during execution of tasks and applications. Figure~\ref{fig:graph_example} shows an example of generated graph. 

\subsection{Internal Communication Bus}

\ac{KWAPI} uses ZeroMQ \cite{HintjensZeroMQ:2013}, a fast broker-less messaging framework written in C++, where transmitters play the role of buffers. ZeroMQ supports a wide range of bus modes, including cross-thread communication, IPC, and TCP. Switching from one mode to another is straightforward. ZeroMQ also provides several design patterns such as publish/subscribe and request/response. As mentioner earlier, in our publish/subscribe architecture drivers are publishers, and data consumers are subscribers. If no data consumer is subscribed to receive data from a given driver, the latter will not send any information through the network.

Moreover, one or more optional forwarders can be installed between drivers and data consumers to minimise network usage. Forwarders are designed to act as especial data consumers who subscribe to receive information from a driver and multicast it to all normal data consumers subscribed to receive the information. Forwarders enable the design of complex topologies and optimisation of network usage when handling data from multiple sites. They can also be used to bypass network isolation problems and perform load balancing.

\subsection{Interface with Ceilometer}

We opted for integrating KWAPI with an existing open source cloud platform to ease deployment and use. Leveraging the capabilities offered OpenStack can help in the adoption of a monitoring system and reduce its learning curve.

Ceilometer's central agent and a dedicated pollster (\textit{i.e.} \ac{KWAPI} Pollster) are used to publish and store energy metrics into Ceilometer's database. They query the REST API data consumer and publish cumulative (kWh) and gauge (W) counters that are not associated with a particular tenant, since a server can host multiple clients simultaneously. 

Depending on the number of monitored devices and the frequency at which measurements are taken, wattmeters can generate a large amount of data thus demanding storage capacity for further processing and analysis. Management systems often store and perform pre-processing locally on monitored nodes, but such an approach can impact on CPU utilisation and influence the power consumption. In addition, resource managers may switch off idle nodes or set them to stand by mode to save energy, which make them unavailable for processing. Centralised storage, on the other hand, allows for faster data access and processing, but can generate more traffic given that measurements need to be continuously transferred over the network to a central point.  

Ceilometer using its own central database, which is used here to store the energy consumption metrics. In this way, systems that interface with OpenStack's Ceilometer, including Nova, can easily retrieve the data. It is important to notice that, even though Ceilometer provides the notion of a central repository for metrics, it also uses a database abstraction that enables the use of distributed systems such as Apache Hadoop HDFS\footnote{http://hadoop.apache.org/}, Apache Cassandra\footnote{http://cassandra.apache.org/}, and MongoDB\footnote{http://www.mongodb.org/}. 

The granularity at which measurements are taken and metrics are computed is another important factor because user needs vary depending on what they wish to evaluate. Taking measurements at one-second interval or smaller is common under several scenarios, which can be a challenge in an infrastructure comprising hundreds or thousands of nodes, demanding efficient and scalable mechanisms for transferring information on power consumption. Hence, in the next section we evaluate the throughput of KWAPI under a few scenarios.

\section{Performance Evaluation}
\label{sec:performance}

This section provides results of a performance evaluation carried out in our testbed. The goal is not to compare publish/subscribe systems since such work has already been performed elsewhere \cite{EugsterSurvey:2003,FabretPS:2001}. The evaluation demonstrates that the framework serves well the needs of a large range of users of the Grid'5000 platform \cite{Grid5000} --- the infrastructure we use and where the Kwapi framework is currently deployed in production mode as the means for collecting and providing energy consumption information to users. 

Firstly, we want to evaluate the CPU and network usage of a typical driver to observe the framework's throughput, since provisioning a large number of resources for monitoring purposes is not desirable. For this experiment we deployed the \ac{KWAPI} drivers and API on a machine with a Core 2 Duo P8770 2.53Ghz processor and 4GB of RAM. We considered:

\begin{itemize}
 \item a scenario where we emulated 1,000 \ac{IPMI} cards, each card monitored by a driver thread placing a measurement per second on the communication bus.
 \item a case with 100 \acp{PDU} with 10 outlets each and each \ac{PDU} monitored by a driver thread placing ten values per second on the bus. 
\end{itemize}

Under both scenarios, 1,000 measurements per second were placed on the bus, even though monitoring was done using different types of probes. We have evaluated these scenarios considering both message signature enabled and disabled. Table~\ref{tab:parameters_usage} summarises the considered scenarios.

\begin{table}
\centering
\caption{Scenarios considered in the experiments.}
\label{tab:parameters_usage}
\begin{tabular}{lcc}
\toprule
\textbf{Scenario name} & \textbf{Agent thread scheme} & \textbf{Message signature}  \\
\toprule
IPMI message signed     & 1 thread per card & Enabled\\
\midrule
IPMI message unsigned   & 1 thread per card & Disabled\\
\midrule
PDU message signed     & 1 thread per PDU & Enabled\\
\midrule
PDU message unsigned   & 1 thread per PDU & Disabled\\
\bottomrule
\end{tabular}
\end{table}

\begin{figure}[!ht]
\center
\includegraphics[width=1.\columnwidth]{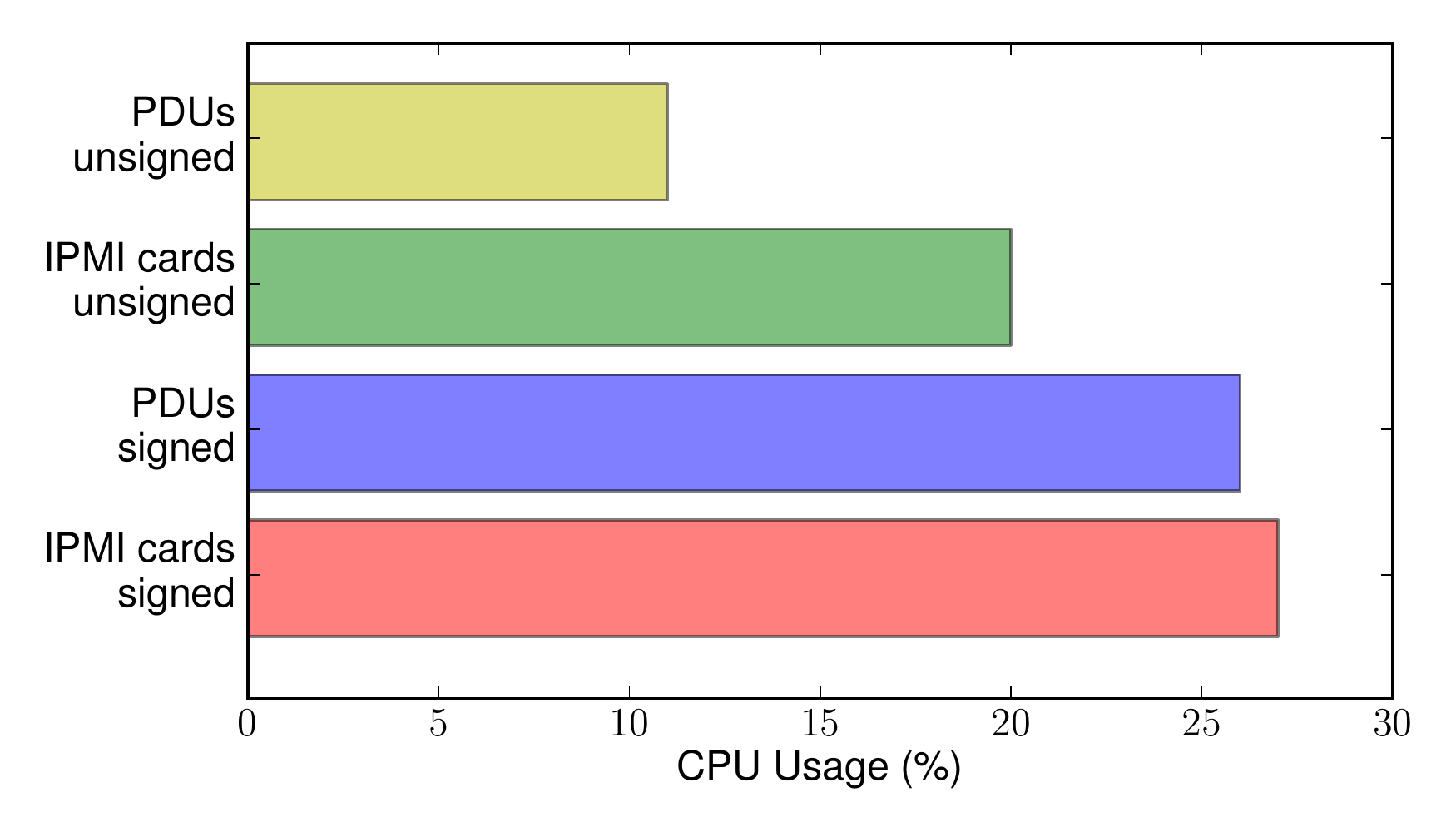}
\caption{Driver CPU usage under the evaluated scenarios.}
\label{fig:cpu_usage}
\end{figure}

Figure~\ref{fig:cpu_usage} shows the results of CPU usage of drivers under the evaluated scenarios. The socket type and number of driver threads do not seem to have a distinguishable impact on the CPU usage. On the test machine, the \ac{KWAPI} drivers with message signature disabled (\textit{i.e.} \ac{IPMI} cards unsigned and \acp{PDU} unsigned) consumed on average 20\% of the total CPU power. 


\begin{figure}[!ht]
\center
\includegraphics[width=1.\columnwidth]{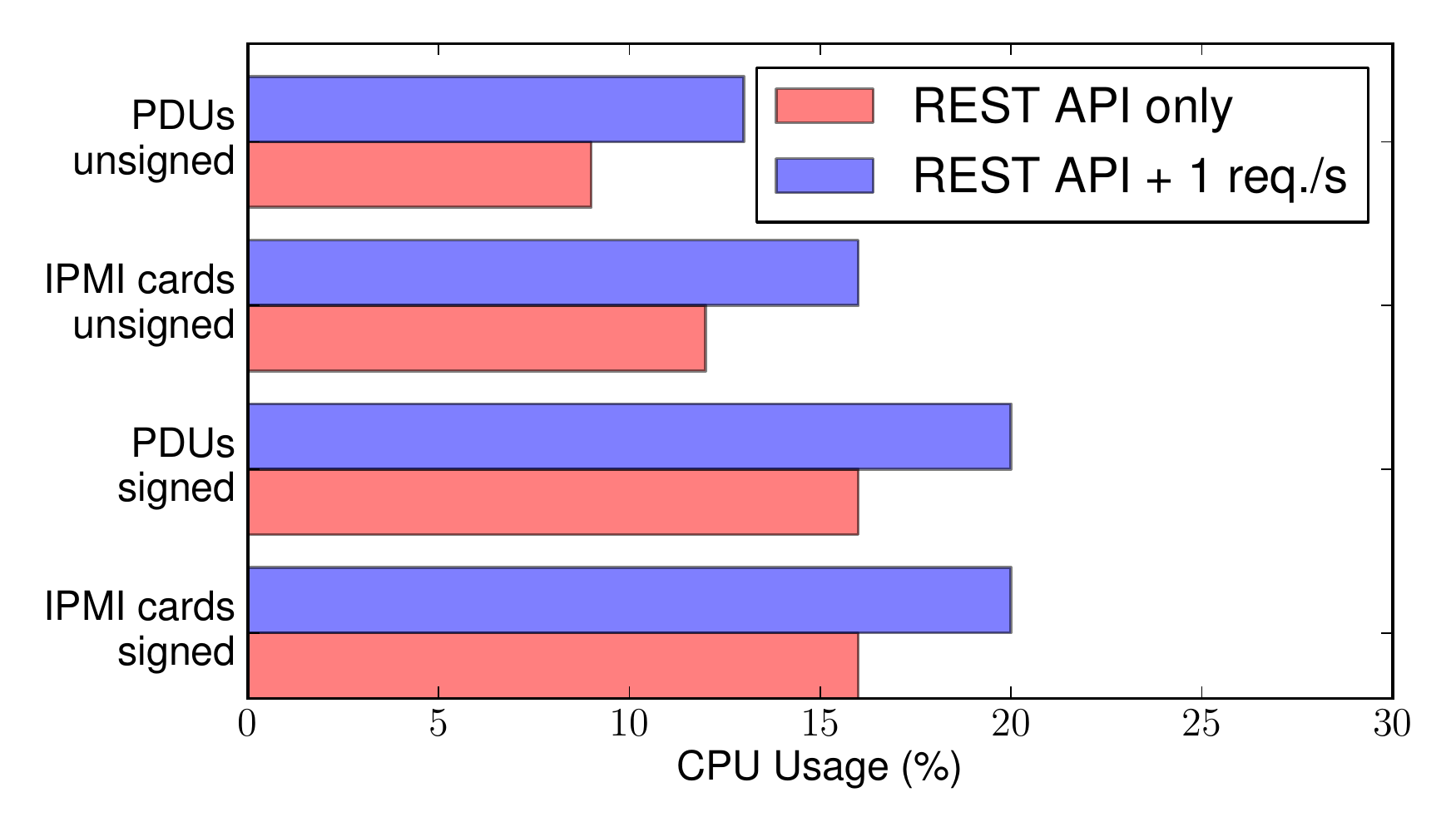}
\caption{API consumer CPU usage under the evaluated scenarios.}
\label{fig:cpu_usage_consumer}
\end{figure}

We also evaluated the CPU consumption of the REST API data consumer under the scenarios described in Table \ref{tab:parameters_usage}. In addition to these scenarios, two conditions were assessed, namely (i) the REST API working as a consumer requesting data from drivers at a one-second time interval (REST API only); and (ii) the API requesting data at one-second interval and also answering a call every second to provide the collected data to an external system (REST API + 1 req/s). Figure \ref{fig:cpu_usage_consumer} summarises the obtained results. The CPU consumption is in general low, and even when message signing is enabled and the API serves a query, its consumption is below 20\%. The small variation between the scenarios without message signing is caused by the manner ZeroMQ accumulates data on nodes prior to transmission. 

Although the CPU usage often depends on the drivers, data consumers, and their complexity, and whether message signature is enabled, the experiments show that a large number of probes can be managed by a single machine. In our environment, a management machine per site is more than enough to accommodate the users' monitoring needs. The drivers and API can reuse a machine that already serves other monitoring purposes.

\begin{figure}[!ht]
\center
\includegraphics[width=1.0\columnwidth]{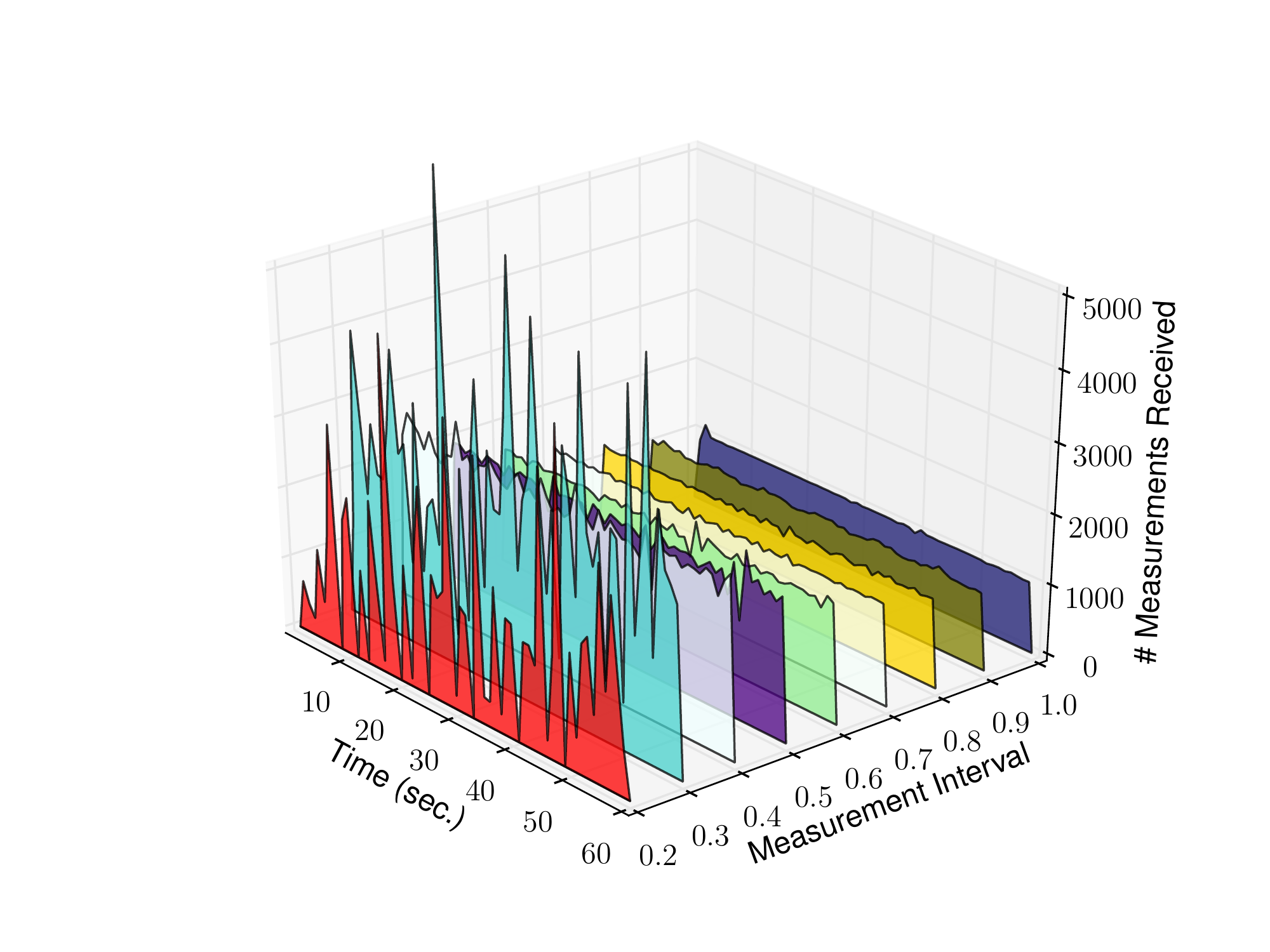}
\caption{Number of observations received over a 60 second interval under different multiple intervals.}
\label{fig:measurement_intervals}
\end{figure}



Although a measurement interval of one second meets the requirements of users in our platform, we wanted to evaluate the impact of using a communication bus in the transfer of observations between drivers and the REST API consumer. In a second experiment we used two machines. On the first machine we instantiated 1,000 driver threads placing random observations on the communication bus. On the second machine we measured the number of measurements that the API is able to receive over a minute. We varied the time between measurements from 0.2 to 1.0 seconds. Figure~\ref{fig:measurement_intervals} summarises the obtained results. Though the number of observations generated in this experiment is much higher than what we currently need to handle in our platform, we observe that the framework is able to transfer measurements from drivers to API under a 0.4 second interval without adding much jitter. Under smaller measurement intervals, however, observations start to accumulate and are transferred at large chunks. We believe that under small measurement intervals, and consequently a very large number of observations per second, an architecture based on stream processing systems that guarantees data processing might be more appropriate. Hence, although the framework suits the purposes of large range of users, if measurements are to be taken at very small time intervals, a stream processing architecture would probably yield better performance by enabling the placement of elements to preprocess data closer to where it is generated.



\section{Conclusion}
\label{sec:conclusion}

In this paper, we described a novel framework (KWAPI) for monitoring the power consumed by resources of an Openstack cloud. Based on lessons learned by monitoring the power consumption of large distributed infrastructure, we proposed an energy monitoring architecture based on a publish/subscribe model. The framework works in tandem with OpenStack's Ceilometer. Experimental results demonstrate that the overhead posed by the monitoring framework is small, allowing us to serve the users' monitoring needs of our large scale infrastructure. 

As future work, we intend to explore means to increase the monitoring granularity and the number of measured devices by applying a hierarchy of plug-ins, and a stream processing system with guarantees on data processing \cite{Storm,S4} for processing streams of measurement tuples.


\section*{Acknowledgments}

This research is supported by the French Fonds national pour la Soci\'{e}t\'{e} Num\'{e}rique (FSN) XLCloud project. Some experiments presented in this paper were carried out using the Grid'5000 experimental testbed, being developed under the Inria ALADDIN development action with support from CNRS, RENATER and several Universities as well as other funding bodies (see https://grid5000.fr). Authors wish to thank Julien Danjou for his help during the integration of \ac{KWAPI} with OpenStack and Ceilometer.

\bibliographystyle{IEEEtran}
\balance
\bibliography{references}

\end{document}